\documentclass[12pt,preprint]{aastex}

\lefthead{Escala}
\righthead{On the functional dependence of the universal star formation law.}

\begin{document}

\title{On the Functional Form of the Universal Star Formation Law.}

\author{Andr\'es Escala}
\affil{Departamento de Astronom\'{\i}a, Universidad de Chile, Casilla 36-D, Santiago, Chile.}
\affil{aescala@das.uchile.cl}

\begin{abstract}
We study the functional form of the star formation law, using the Vaschy-Buckingham Pi theorem. We find that that it should have a form $\rm  \dot{\Sigma}_{\star} \propto \sqrt{\frac{G}{L}}\Sigma_{gas}^{3/2}$, where L is  a characteristic length that is  related with an integration scale. With a reasonable estimation for L, we find that galaxies from different types and redshifts, including  Low Surface Brightness galaxies, and individual star-forming regions in our galaxy, obey this single star formation law. We also find that depending on the assumption  for L, this   star formation law adopt     different formulations of $\rm  \dot{\Sigma}_{\star}$  scaling, that are widely studied in the literature:   $\rm  \Sigma_{gas}^{3/2}, \Sigma_{gas}/t_{orb}, \Sigma_{gas}/t_{ff} \, and \,  \Sigma_{gas}^{2}/v_{turb}$. We also  study secondary controlling parameters of the star formation law, based on the current evidence from numerical simulations and find that for galaxies, the star formation efficiency   should be controlled, at least, by the turbulent Toomre parameter, the sonic and Alfv\'enic Mach numbers.

\end{abstract}


\section{Introduction}

Galaxies are building blocks of the universe,  the galaxies themselves are constituted by stars, therefore to understand  the rate that galaxies forms their stars is a fundamental part in the our understanding of how the universe  evolves. For that reason,  considerable effort has been performed to understand the rate at which galaxies fills  the cosmos with stars.

Observations  of normal spiral galaxies by Schmidt (1959)   originally
suggested  that their star formation rates (SFRs) scale with their gas content. This  was extended to  galaxies with higher SFR  by Kennicutt (1998), leading to an empirical law for star formation called the Kennicutt-Schmidt (KS) Law: $ \rm \dot{\Sigma}_{star} =  \epsilon_{SF}\Sigma_{gas}^{1.4} \,\,\, ,$ where $\rm  \Sigma_{gas}$ and $\rm \dot{\Sigma}_{star}$ are the gas surface density and SFR per unit area. However, Bigiel et al. (2008), Leroy et al. (2008), Wyder et al. (2009) and Shi et al. (2011) found deviations from the $\sim$ 1.4 slope at lower surface densities. In addition, Daddi et al. (2010b) and Genzel et al. (2010) studied  this relation for  high redshift galaxies, Shi et al. (2014),   the effects of metallicity on the SFR and  Guillard et al. (2014) the role of radio jets, all finding possible departures from a single law. Also, for major mergers Xu et al. (2014) and Hodge et al. (2015)  found higher SFRs  for spatially resolved individual regions.

On the theoretical side, considerable  literature focus on   analytic calculations with a considerable level of assumptions and free parameters, being hard to test their  validity  against observed data (Krumholz \& McKee 2005, Krumholz et al. 2012, Hopkins 2013, just to mention  few attempts). Moreover, several galactic-scale numerical simulations (Li, Mac Low \& Klessen 2005, Tasker \& Bryan 2006, Stinson et al. 2006, Tasker \& Tan 2009, Becerra \& Escala 2014), using completely different thermal physics, accuracy of the hydrodynamic method, star formation/feedback prescriptions, etc., are all able to  find SFR in  agreement with the KS Law, regardless of the different physics implemented. 

In this paper we propose a different approach, to use Vaschy-Buckingham  theorem to  guide   the analysis of the current observational and numerical evidence in the subject, to see what we can learn from them and if it is possible, to infer  functional forms  and controlling parameters of the universal star formation law. Our goal is to summarize the current evidence into a unique physical equation valid at all scales, in which variations of its physical variables, explain the  variations of the observed star formation rates from  individual clouds in the Milky Way, to galaxies in the early universe.

This work is organized as follows. We start with a dimensional analysis of the star formation law,
  in order to find a  physical relation in agreement with the current observational and numerical evidence in \S 2. Section 3 continues with a discussion on the characteristic length introduced in \S 2, and test  candidates against  the star formation rates in  galaxies of different types and redshifts. In \S 4  we study the physics that determines this characteristic length, deriving 
  several formulations for the star formation law that appears in the literature. Finally in \S 5, we discuss  the  results of this work.

\section{Dimensional Analysis of the Star Formation Law}

The Vaschy-Buckingham Pi theorem defines the rules to be fulfilled by any meaningful physical relation and it is a formalization of Rayleigh's method of dimensional 
analysis. The theorem states that if there is a physically meaningful equation involving a certain number, n, of physical variables, and k is the number of 
relevant dimensions, then the original expression is equivalent to an equation involving a set of p = n $-$ k  dimensionless parameters constructed from the 
original variables. Mathematically speaking, if we have the following physical equation:
\begin{equation}
F(A_1,A_2,\ldots,A_n)=0 \, ,
\label{}
\end{equation} 
where the Ai are the n physical variables that  are expressed in terms of k independent physical units, Eq 1 can be written as
\begin{equation}
f(\Pi_1, \Pi_2, \ldots, \Pi_{n-k})=0\,  ,
\label{}
\end{equation} 
where the $\rm \Pi_{i}$ are dimensionless parameters constructed from the $\rm A_i$, by p = n $-$ k dimensionless equations  of the form $\rm \Pi_i=A_1^{m_1}\,A_2^{m_2}\cdots A_n^{m_n}$. 
We will use this theorem to design a physically meaningful equation for the universal star formation law.

In the process of finding  the functional form of any physical equation thru dimensional analysis, is critical to choose the relevant physical variables. We will perform this iteratively for the universal star formation law. Since this problem has a minimum of 3 relevant dimensions, mass [M], length [L] and time [T], we need at least four physical variables in order to have one dimensionless parameter $\rm \Pi_{1}$. Motivated by the Kennicutt-Schmidt  Law, $\rm  \dot{\Sigma}_{\star}$ and $\rm \Sigma_{gas}$ must be the two first physical variables. There is little doubt that gravity plays a critical role in the star formation problem, ever since the early analytical results by Jeans (1902), Bonnor (1956) and Ebert (1955), and numerical work by Larson (1969) and Penston (1969). Therefore, G should appear somewhere in any star formation law and is our third choice. For our final fourth physical variable, both a characteristic time or length could work, however, to avoid the trivial dependence proportional to  $\rm  \Sigma_{gas}/\tau$ (e.g. Silk 1997, Elmegreen 1997), we choose to use a characteristic length that we will called it L. 

To find the first dimensionless parameter  is straightforward by looking integer exponents such $\rm \Pi_{1} = G^{a_1} \Sigma_{gas}^{b_1} L^{c_1} \dot{\Sigma}_{\star}$ has no dimensions. This is equivalent to force $\rm [L]^{3a_1-2b_1+c_1-2}$ $\rm [M]^{-a_1+b_1+1}$ $\rm [T]^{-2a_1-1}$ to  be dimensionless,  which has the unique solution of $\rm a_1$=-1/2, $\rm b_1$=-3/2 and $\rm c_1$=1/2. This implies that the dimensionless parameter is $\rm \Pi_{1} = G^{-1/2} \Sigma_{gas}^{-3/2} L^{1/2} \dot{\Sigma}_{\star}$ and if the star formation law depends only on this four physical variables, the Vaschy-Buckingham Pi theorem states that should be  a function f such $\rm f(\Pi_1=G^{-1/2} \Sigma_{gas}^{-3/2} L^{1/2} \dot{\Sigma}_{\star})=0$. Additionally, if f have a zero that we called $\rm \epsilon$, such f($\rm \epsilon$)=0, this implies :

\begin{equation}
\rm  \dot{\Sigma}_{\star} = \epsilon \sqrt{\frac{G}{L}}\Sigma_{gas}^{3/2}\,
\label{elaw1}
\end{equation} 
which has a dependence on surface density quite similar to the standard formulation for the star formation law, but has an extra term L. If such characteristic length is constant, we recover the KS  Law with almost the observed slope (n$\sim$1.4; Kennicutt 1998). In summary, the Vaschy-Buckingham Pi theorem is telling us that the standard formulation of the star formation law needs to be at least corrected by a characteristic length, in order to have the proper dimensions. 

In addition to this length correction, other physical variables should have a role in controlling the rate at which galaxies forms their stars in the universe. We will explore few possibilities subsequently.

\subsection{Role of turbulence}

At intermediate, Giant Molecular Cloud (GMC) scales, is commonly believed that turbulence governs the GMCs dynamics, with typical thermal Mach numbers $(\rm  \mathcal{M}_s=v_{turb}/C_{S})$ are of the order 10-20 (Mac Low \& Klessen 2004). Moreover, as it has been suggested by Larson (1979, 1981) and others (see for example Ballesteros-Paredes et al. 2007 and references therein), that the structure and dynamics of the ISM on these intermediate scales is roughly self-similar and described by power laws, as in a turbulent cascade.

In order to explore if the three dimensional rms velocity of turbulent motions, $\rm v_{turb}$, is a fifth  physical variable in the star formation law, a $\rm \Pi_{2}$ should be  constructed since Vaschy-Buckingham Pi theorem allows now 5-3=2  dimensionless parameters. The parameter $\rm \Pi_{2} =  v_{turb} G^{a_2} \Sigma_{gas}^{b_2} L^{c_2}$ has no dimensions for a unique solution of $\rm a_2$=-1/2, $\rm b_2=$-1/2 and $\rm c_2=$-1/2, implying a second dimensionless parameter  $\rm v_{turb} G^{-1/2} \Sigma_{gas}^{-1/2} L^{-1/2}$. In this case, the Pi theorem states that there is an equation  $\rm f(\Pi_1,\Pi_2)=0$ and if f is regular and differentiable, we can use the  implicit function theorem to advocate the existence of a function $\Pi_{1}=\epsilon(\Pi_{2})$. This implies that if the turbulent rms velocity is an additional  physical variable in the star formation law, Eq 3 should be be replaced by:

\begin{equation}
\rm  \dot{\Sigma}_{\star} = \epsilon[\frac{v_{turb}}{\sqrt{G \Sigma_{gas} L}}] \sqrt{\frac{G}{L}}\Sigma_{gas}^{3/2}\, 
\label{elaw2}
\end{equation} 
where $\rm \epsilon$ is now a function of $\rm v_{turb} (G \Sigma_{gas} L)^{-1/2}$, a parameter that quantifies the relative strength of turbulence and gravity. 

The  dependence of $\rm \epsilon$  on the second dimensionless parameter $\rm \Pi_2$, can be directly compared against numerical experiments that  studies fragmentation in turbulent GMCs. Padoan et al. (2012) found that the SFR per free fall time, $\rm \epsilon_{ff}$, strongly depends on the free-fall time per turbulent crossing time. The turbulent crossing time  is defined by Padoan et al. (2012) as $\rm t_{dyn}=L/2v_{turb}$ and the free-fall time as $\rm t_{ff}= (3\pi/32G\rho_{0})^{1/2}$, where $\rm \rho_{0}$ is the mean density. For $\rm \rho_{0} = \Sigma_{gas}/L$, is straightforward to find that their Eq 1 is equivalent to our Eq 4, in particular, that $\rm \epsilon_{ff}$ has the same dimensionless dependence  that $\rm \epsilon(\Pi_{2})$, within geometrical factors ($\rm t_{ff}/t_{dyn} = (3\pi/8)^{1/2} v_{turb} (G \Sigma_{gas} L)^{-1/2}$). 

Using the Pi theorem, we also found the main dependence suggested by state-of-art simulations of turbulent fragmentation, since this dimensionless parameter $\rm \Pi_{2}$, is also equivalent to the virial parameter, $\rm \alpha_{virial}=2T/W$, used in previous works (Padoan \& Nordlund 2011). Padoan et al. (2012) quantified this strong dependence by the fitting formula $\rm \epsilon_{ff} \propto exp(-1.6 t_{ff}/t_{dyn})$, which is equivalent to  $\rm \epsilon_{ff} \propto exp(-1.74 v_{turb} (G \Sigma_{gas} L)^{-1/2})$.


\subsection{Small scale physics: role of thermal pressure and magnetic fields}

There is little doubt, that final barrier that should overcome the self-gravity of interstellar gas to form a star, is thermal pressure. In addition, magnetic fields are often advocated as a relevant source of support (Mouschovias 1974, Shu 1977). We will start analyzing the role of thermal pressure, since is what eventually stops the collapse at stellar densities.

To include the thermal sound speed, $\rm   C_{S}$, as the sixth physical variable in the star formation law, a  $\rm \Pi_{3}$ should be constructed. The parameter $\rm \Pi_{3} =  C_{S} G^{a_3} \Sigma_{gas}^{b_3} L^{c_3}v_{turb}^{d_3}$ has no dimensions for two possible solutions: i) $\rm a_3=b_3=c_3=0$, $\rm d_3=-1$ ii) $\rm a_3=b_3=c_3=-1/2$, $\rm d_3=0$. We will focus on the case i),  $\rm \Pi_{3}=C_{S}/v_{turb}$, since there is numerical work that studies the role of the sonic mach number $\rm  \mathcal{M}_s=v_{turb}/C_{S}$.

Using MHD simulations, covering a substantial range of observed  cloud parameters with Mach numbers  $\rm  \mathcal{M}_s=v_{turb}/C_{S}= 5-50$, Federrath (2013) found that the observed scatter in the star formation law, can be primarily explained by physical variations in the turbulent Mach number  $\rm  \mathcal{M}_s$. This work also found that magnetic fields reduce the star formation efficiency, $\rm \epsilon$, but only very marginally.

To include  magnetic fields in our analysis, is easier using the Alfv\'enic  velocity, $\rm v_a$, in order to compare their   strenght relative to  thermal pressure. A   $\rm \Pi_{4}=  v_{a} G^{a_4} \Sigma_{gas}^{b_4} L^{c_4}v_{turb}^{d_4}$ should be constructed, that has again no dimensions for two possible solutions: i) $\rm a_4=b_4=c_4=0$, $\rm d_4=-1$ ii) $\rm a_4=b_4=c_4=-1/2$, $\rm d_4=0$. For the first case,  $\rm \Pi_{4}=v_{a}/v_{turb} = \mathcal{M}_a^{-1}$, is the inverse of the Alfv\'enic Mach number.

Padoan et al. (2012) found that the star formation efficiency, $\rm \epsilon$, has a complex but weak dependence  on $\rm \mathcal{M}_a$, varying by less than a factor of two for characteristic  values of $\rm \mathcal{M}_a$. However, this work found that $\rm \epsilon$ is insensitive to variations of $\rm  \mathcal{M}_s$. This disagreement with Federrath (2013), might be due to the dynamical range studied: $\rm  \mathcal{M}_s=10-20$ in  Padoan et al. (2012), compared to  $\rm  \mathcal{M}_s= 5-50$ in Federrath (2013). The range $\rm  \mathcal{M}_s=10-20$ might be justified for local GMCs, but for extragalactic sources such starbursts or high z galaxies, a range like the one in Federrath (2013) is better justified.

Including thermal pressure and magnetic fields, the Pi theorem states that there is an equation  $\rm f(\Pi_1,\Pi_2, \Pi_3,\Pi_4)=0$ and if f is regular and differentiable, the implicit function theorem guarantee  the existence of a function  $\Pi_{1}=\epsilon(\Pi_{2}, \Pi_3,\Pi_4)$. The latter is equivalent to a star formation law of the form:

\begin{equation}
\rm  \dot{\Sigma}_{\star} = \epsilon[\frac{v_{turb}}{\sqrt{G \Sigma_{gas} L}}, \mathcal{M}_s,\mathcal{M}_a] \sqrt{\frac{G}{L}}\Sigma_{gas}^{3/2}\, .
\label{elaw3}
\end{equation}

The  options ii) for $\rm \Pi_3 \,\, and \,\, \Pi_4$ are respectively, $\rm C_{S} (G \Sigma_{gas} L)^{-1/2}$ and $\rm v_{a} (G \Sigma_{gas} L)^{-1/2}$. These options are equivalent to $\rm \Pi_3 = \Pi_2/\mathcal{M}_s$ and $\rm \Pi_4 = \Pi_2/\mathcal{M}_a$, a combination of the previous dimensionless parameters.

We can continue with this iterative process of searching for physical variables, by including other controlling parameters suggested in the literature, such as the molecular mass fraction $\rm f_{H_2}$  (Krumholz \& McKee 2005) or the metalicity (Dib et al. 2011, Shi et al. 2014). However, we prefer to  focus in this paper on the dynamical  variables just mentioned, that are motivated by both observational and numerical studies.

\section{On the Characteristic Length L}

We have been able to find a star formation law that, in addition to be dimensionally correct, have several dependences in agreement with both the KS Law (for a constant characteristic length L) and numerical experiments that studies  star formation within GMCs. However, so far characteristic length L is a free parameter without any physical interpretation.

We will start with the simplest possible choice for characteristic length L: the region total  radius R, which is a natural scale of galaxies and star-forming regions, to later evolve to more sophisticated choices.  Since the goal is to find a universal law valid at all scales, we include data from individual star forming clouds, up to extended galaxies like Low Surface Brightness (LSB) galaxies (five order of magnitude variations in R). We also include normal spiral, local starburst and high redshift galaxies.

In the large dynamical range studied, that needs to be displayed on a  log-log plot, the relation is   dominated by order of magnitude variations of the primary dependences. For that reason, $\rm \Pi_{1}$ will dominate over the other dependences, therefore, for simplicity we will assume $\epsilon(\Pi_{2},\Pi_{3}, \Pi_4)=\epsilon$,  constant during this section (Eq. \ref{elaw1}).

Figure 1 shows the SFR per unit area, against $\rm \Sigma_{gas}R^{-1/3}$, being R the  radius  of each galaxy or star-forming regions. The gas surface density $\rm \Sigma_{gas}$ and SFR per unit area $\rm  \dot{\Sigma}_{\star}$, for each data point was taken for   Low Surface Brightness (LSB) galaxies (Wyder et al. 2009), normal spirals/local starbursts (Kennicutt 1998), high redshift disks (Daddi et al. 2010a; Tacconi et al. 2010), high redshift starbursts (Genzel et al. 2010) and galactic GMCs (Lada et al. 2010, Hierderman et al. 2010).  In addition to the previous references, for the radius R,  data was taken from  Young et al. (1995) for normal spirals, Smith \& Harvey (1996), Downes \& Solomon (1998), Kenney et al. (1992), Wild et al. (1992) and Telesco et al. (1993) for local starbursts,  Genzel et al. (2010)  for high redshift disks and Krumholz et al. (2012) for galactic GMCs. 

\begin{figure}[h!]
\begin{center}
 \includegraphics[width=11.9cm]{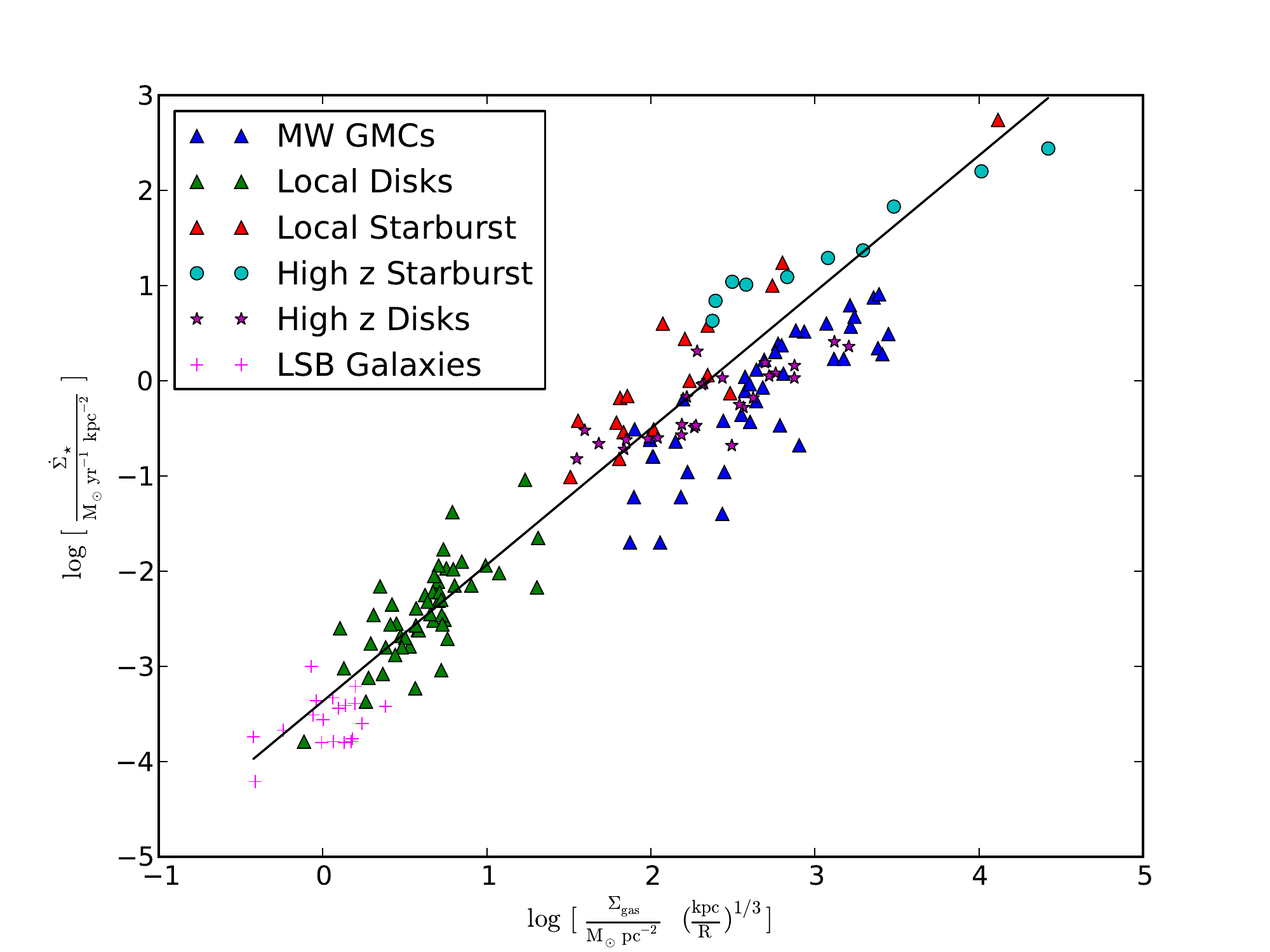}
\caption{ SFR density as a function of the gas  surface density, divided by the 1/3 power of the radius. The symbols displayed  are: Blue triangles are galactic GMCs taken from Lada et al. (2010) and Hierderman et al. (2010). Green and red triangles are   local spiral galaxies and (U)LIRGs from Kennicutt (1998). The purple stars and cyan filled circles  are high z disks (Daddi et al. 2010a; Tacconi et al. 2010) and starburst galaxies (Genzel et al. 2010).  Pink crosses are LSB galaxies from Wyder et al. (2009).}
\label{F1}
\end{center}
\end{figure}

All  galaxy types displayed in Fig 1, including local and high z galactic disks, starbursts at different redshifts and even LSB galaxies  are  described by a single relation with  slope consistent with 1.5, the one expected from Eq \ref{elaw1}. The  black curve in Fig \ref{F1}, is  the best fit to the sample of different galaxy populations (not including individual star-forming clouds), which corresponds to a slope of 1.43. Surprisingly, even individual star-forming clouds follows a similar trend, the expected from Eq \ref{elaw1}, but shifted towards lower values of SFR per unit area. These results suggest that we are on the right track, but there is a fundamental difference in L between galaxies and  star-forming clouds.

Is important to notice,  besides that the galactic radius R is not (a priori) a particularly meaningful scale in the star formation problem, the  $\rm R^{-1/2}$ term in Eq \ref{elaw1} is already able to erase the difference in star formation efficiency between spirals and LSB galaxies, for which considerable literature has been written  (Bigiel et al. 2008, 2010; Wyder et al. 2009; Shi et al. 2011). Because  LSB are typically more extended than normal spirals,  almost a factor 10 in the more extreme cases of Malin 1 or LSBC F568-06, the decrease due to the factor  $\rm R^{-1/2}$  explains  their lower SFR per unit area for a fixed $\rm \Sigma_{gas}$. Therefore, a term that scales similar to  R will be a good candidate for characteristic length L.

Star formation is inherently a three dimensional problem and  the star formation law, is expressed in terms of some two dimensional quantities: $\rm  \dot{\Sigma}_{\star}$ and $\rm \Sigma_{gas}$. The length-scale responsible for such transition in dimensionality, an integration, is a natural candidate for being the  characteristic length L. From an observational point of view, this integration scale will be in the observer's line of sight (LOS). In 3-D numerical simulations,  will be the one chosen by the theorist, which  is in most cases  the vertical scale length. The difference between both cases is  projection factor, unless we are dealing with an edge-on disk, that we will neglect   in this section and leave  it into the scatter, thus, we will focus on estimations of the vertical scale length. We will come back to this point in \S3.2

We will  now estimate the characteristic length L to be equal to the vertical scale length, $\rm H=\eta R$, and to avoid  ad-hock fine tunning, we will distinguish only between galaxy/regions  types. For typical disk galaxies,  H/R is typically a few percent, therefore  $\eta$=0.02 is a good choice for LSB galaxies and normal spirals. Nuclear disks of starburst galaxies are much turbulent and thicker (Downes \& Solomon 1998) and $\eta$=0.1 will be our choice in such case. Similar case are high-z disks and starbursts, justifying to choose again $\eta$=0.1. Finally, since galactic GMCs are roughly spherical, in this case $\eta$=2.

\begin{figure}[h!]
\begin{center}
\includegraphics[width=11.9cm]{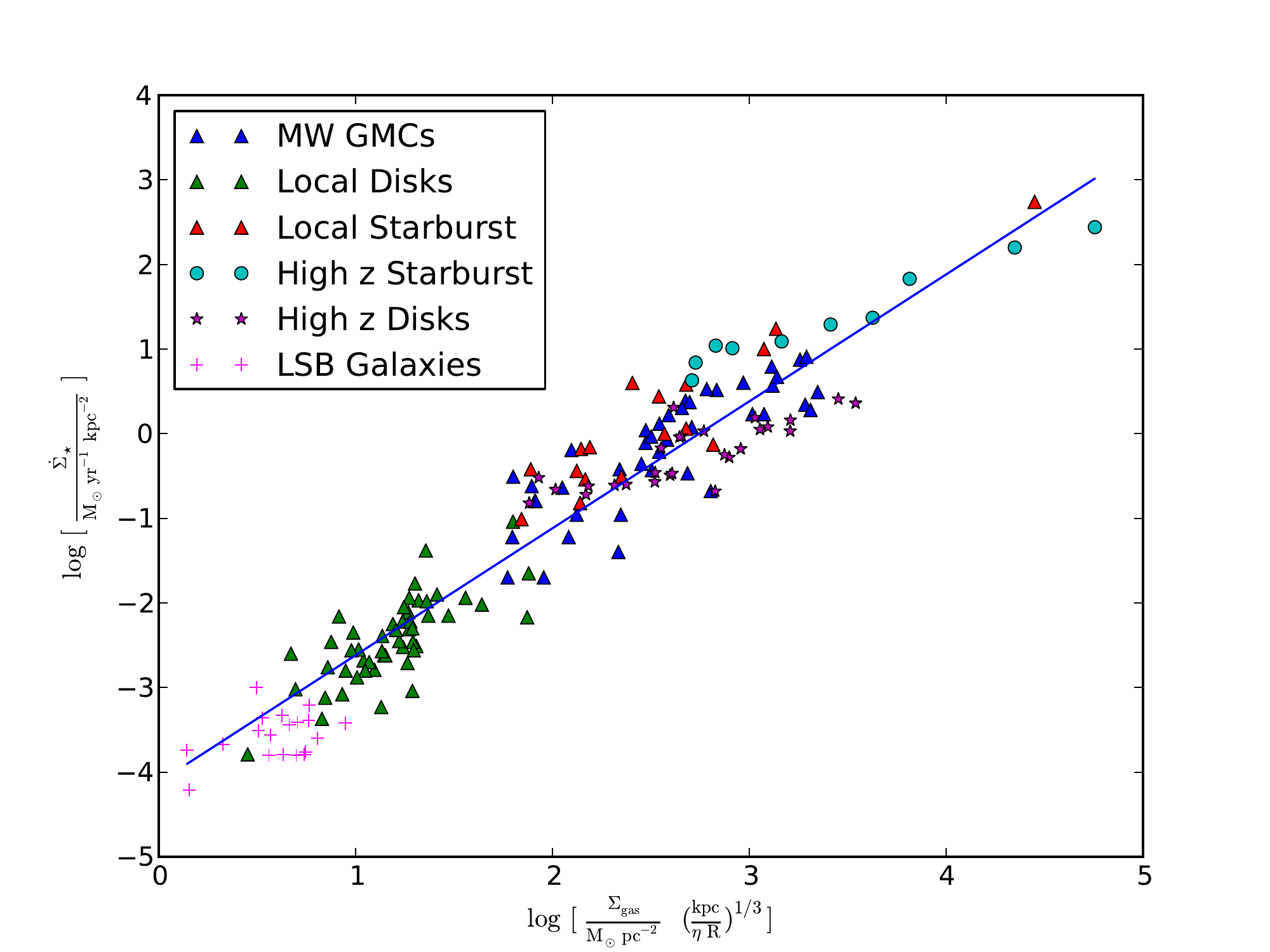}
\caption{Same as Figure \ref{F1}, but with the gas surface densities divided by  the 1/3 power of the vertical scale length ($\rm H=\eta R$). The blue line shows the fitting relation  given in Equation \ref{fit}, that has a scatter of 0.43 dex.}
\label{F2}
\end{center}
\end{figure}
   
Figure \ref{F2} shows the SFR per unit area, against $\rm  \Sigma_{gas}(\eta R)^{-1/3}$, showing that with this simple and more meaningful estimation  for L, all the star forming regions, from local GMCs to high redshift galaxies,  belongs to a single relation of slope 1.5. The blue solid line corresponds to
\begin{equation}
\rm  log\frac{\dot{\Sigma}_{\star}}{M_{\odot}kpc^{-2}yr^{-1}} =  \frac{3}{2}\, (log\frac{\Sigma_{gas}}{M_{\odot}pc^{-2}} - \frac{1}{3} log\frac{\eta R}{kpc}) - 4.1 =    \frac{3}{2}\, log\frac{\Sigma_{gas}}{M_{\odot}pc^{-2}}-\frac{1}{2}log \frac{H}{kpc} - 4.1  \, , 
\label{fit}
\end{equation}
which is the same functional form  of Eq. \ref{elaw1} with $\rm  log(\sqrt G \epsilon)= -4.1$ and has a scatter of 0.43 dex, with respect to the data points. We have chosen to fit the data points with a fixed 3/2 slope, instead  of using the  best fit one, which has a slope of 1.56 and a scatter of 0.429 dex, in order to preserve the correct dimensionality inferred from Pi theorem. Deviations from the functional form with the correct dimensionality,  should be associated to new physical parameters, like the ones introduced in \S2.1 and \S2.2, or to variations of intrinsic observational biases.

We found that, thanks to   $\rm H^{-1/2}$ term from Eq. \ref{elaw1}, the data points in Figure \ref{F2}  are consistent with a single the star formation law, with a scatter comparable to the one found by  Daddi et al. (2010b)  for $\rm  \dot{\Sigma}_{\star} \propto \Sigma_{gas}/t_{orb}$. However, the relation fitted by Daddi is not consistent with galactic GMCs (Krumholz et al. 2012), as it is  for our Eq. \ref{elaw1}, that in addition, is consistent with LSB galaxies. Also, Shi et al. (2011) previously  brought  the LSB galaxies to the spiral trend by introducing  a  $\rm \Sigma_{star}^{1/2}$ term, which for a stellar dominated disk potential, with a given stellar velocity  dispersion, is equivalent to a $\rm H^{-1/2}$ term (van der Kruit 1988). It is important to emphasize, that we choose $\rm \eta$ to be in agreement to the observed differences in thickness between the galaxies/star forming regions displayed in Fig \ref{F2}, and is not an ad-hoc extra free parameter  introduced to decrease the scatter in the relation.

The relatively higher scatter seen individual star forming regions, should be expected since at those  smaller scales   other  issues  appears, such  as time sampling in  sub-galactic regions, which are  not included in Eq. \ref{elaw1}. For example, galaxies homogeneously sample the time-line of star formation, whereas individual star-forming regions are at a specific point of such time-line (Kruijssen \& Longmore 2014). Formulations including $\Pi_2$  (Eq. \ref{elaw2}), that can be expressed as the virial parameter, might take into account some of the differences in the evolutionary sequence:  initial collapsing condensations (low $\rm \alpha_{virial}$), steady state configurations ($\rm \alpha_{virial} \sim 1$) and the final evaporation ($\rm \alpha_{virial} >> 1$) . However, for time sampling on even shorter timescales, new parameters should be added.

If the scatter in the relation for  individual clouds is considerably reduced by including $\rm \alpha_{virial}$ or other physical parameters, we could expect those regions depart from the same single relation of galactic systems. This is because the observed variables in star formation laws are always averages at galactic scales, of quantities that vary strongly on the small length scales  of individual clouds. Therefore,  filling factors of gaseous clouds are very different compared to the averaged ones at galactic scales and  this  have an effect in the normalization of the relation. In \S 3.2, we see an example of how the gas surface density and star formation rate are diluted by averaging over larger scales ($\sim$ kpc).

The simplest possible interpretation for the relation showed in Fig \ref{F2} (Eq. \ref{fit}), is in terms of the average free-fall time ($\rm t_{ff}$) at the  characteristic length, which is now $\rm H=\eta R$. Assuming a linear relation between total quantities, $\rm \dot{M}_{\star} = \epsilon M_{gas}/t_{ff}$, dividing it by the total system area $\rm A= R^2$ and noticing that $\rm t_{ff}=1/\sqrt{G\rho}=1/\sqrt{G\Sigma_{gas}/H}$, is straightforward to get $\rm  \dot{\Sigma}_{\star} = \epsilon \sqrt{G/H}\Sigma_{gas}^{3/2}$. However, is important to realize that the agreement between our dimensional analysis and the observed data, is only telling us that such  free fall time is a characteristic timescale of the problem, but this doesn't mean that this simple picture, monolithic free-fall collapse from the  characteristic length, is how it happens in nature. 

The star formation problem is controlled by non-linear physics coming from gravity, turbulence, feedback from stars, etc, that is far more complex than the simple  free-fall interpretation in terms of averaged quantities. Only performing detailed numerical experiments, that includes the relevant physics, could shed light on the exact reasons why this timescale is important. This is analogous to the case of $\rm \Pi_2$, that we were able to find it as relevant parameter using dimensional analysis, but its exponential functional form and reasons for it, arises  from numerical experiments.

Figure \ref{F2} also shows that possible changes in slope in the KS Law at lower surface densities (Bigiel et al. 2008, Wyder et al. 2009), are most probably due to variations of  the  integration scale L, instead of  variations of the molecular gas fraction as proposed by  Bigiel et al. (2008). Moreover, Bigiel et al. (2010) found that the star formation efficiency ($\rm \Sigma_{SFR}/\Sigma_{gas}$), increases with $\rm \Sigma_{gas}$ (dominated by HI in their sample) and decreases with galactocentric radius R. In fact Eq. \ref{elaw1}
 and $\rm L = H=\eta R$ implies a star formation efficiency with the same trend, proportional to $\rm \sqrt{\Sigma_{gas}/ R}$

\subsection{High redshift galaxies}

The high redshift galaxies (filled circles and stars in Figure \ref{F2}), as individual  galaxy populations show  slopes clearly departing from $\sim$3/2.  This arises the question if new physical  parameters are responsible for this  change in slope, as it  was probably the case in the KS Law at low surface densities (the extra $\rm \sqrt{L}$ term).

\begin{figure}[h!]
\begin{center}
\includegraphics[width=11.9cm]{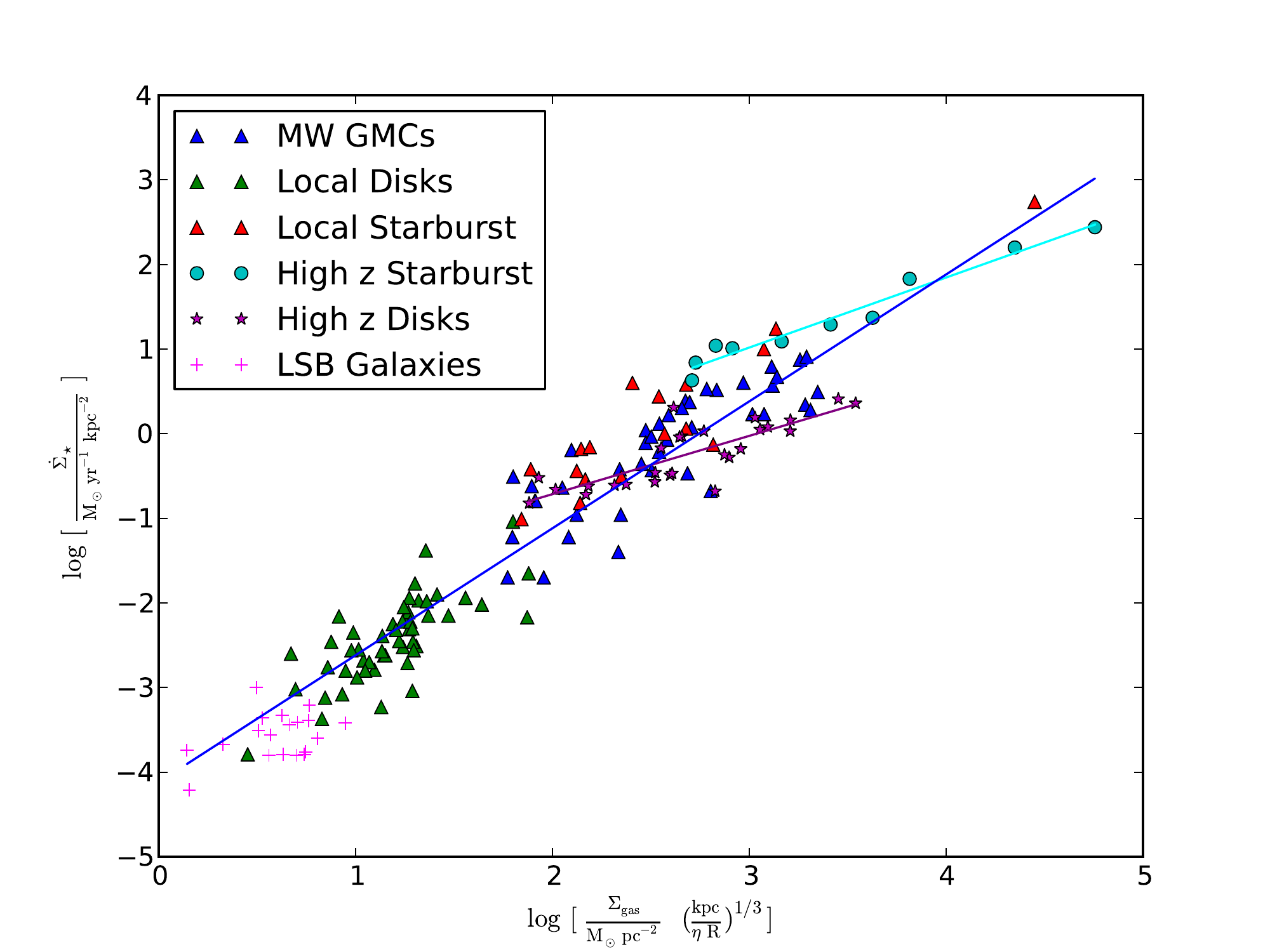}
\caption{Same as Figure \ref{F2}. The purple line shows a slope of $\sim$0.7 for the  individual population of the high z disks (Daddi et al. 2010a; Tacconi et al. 2010). The cyan line shows the slope ($\sim$0.8) for  the population  high z  starburst galaxies (Genzel et al. 2010)}
\label{F3}
\end{center}
\end{figure}

Figure \ref{F3} shows the same data points as Figure \ref{F2}, but with the individual slopes of the high z disks (purple line) and starbursts (cyan line). Both populations have a slope considerably smaller, but surprisingly, quite similar ($\sim$0.7 for disks and $\sim$0.8 for starburst). Since they are high redshift systems, biases and resolution issues are always a possibility, however, variations of other physical  parameters might also produce noticeable changes of the efficiency $\rm \epsilon$. 

In \S 4.1, we will explore the possibility of having the Toomre Q parameter (Toomre 1964), as a physical  parameter controlling the the efficiency $\rm (\epsilon_{SF} \, \propto e^{-Q})$. This allows  to interpret the weaker slope at high z, as a sign of changes in the efficiency. One possible scenario, is that   the high-z galaxies richer in gas are undergoing a more extreme starburst episode, that generates enough energy to be in a state of larger Q, producing a significant decrease of the efficiency that compensates their relatively higher surface densities.

\subsection{The central molecular zone}

So far, we have not distinguished between L being the observer's  LOS or the vertical thickness H, leaving  the difference between both cases into the scatter. This difference is only projection factor, square rooted, unless we are dealing with an edge-on system, in which L should be related to the disk size. For that reason, a  good starting point is to study this effect is the central molecular zone (CMZ), since is nearest edge-on system and that has a a SFR lower than expected from the K-S Law (Kruijssen et al 2014 and references therein).
 
\begin{figure}[h!]
\begin{center}
\includegraphics[width=11.9cm]{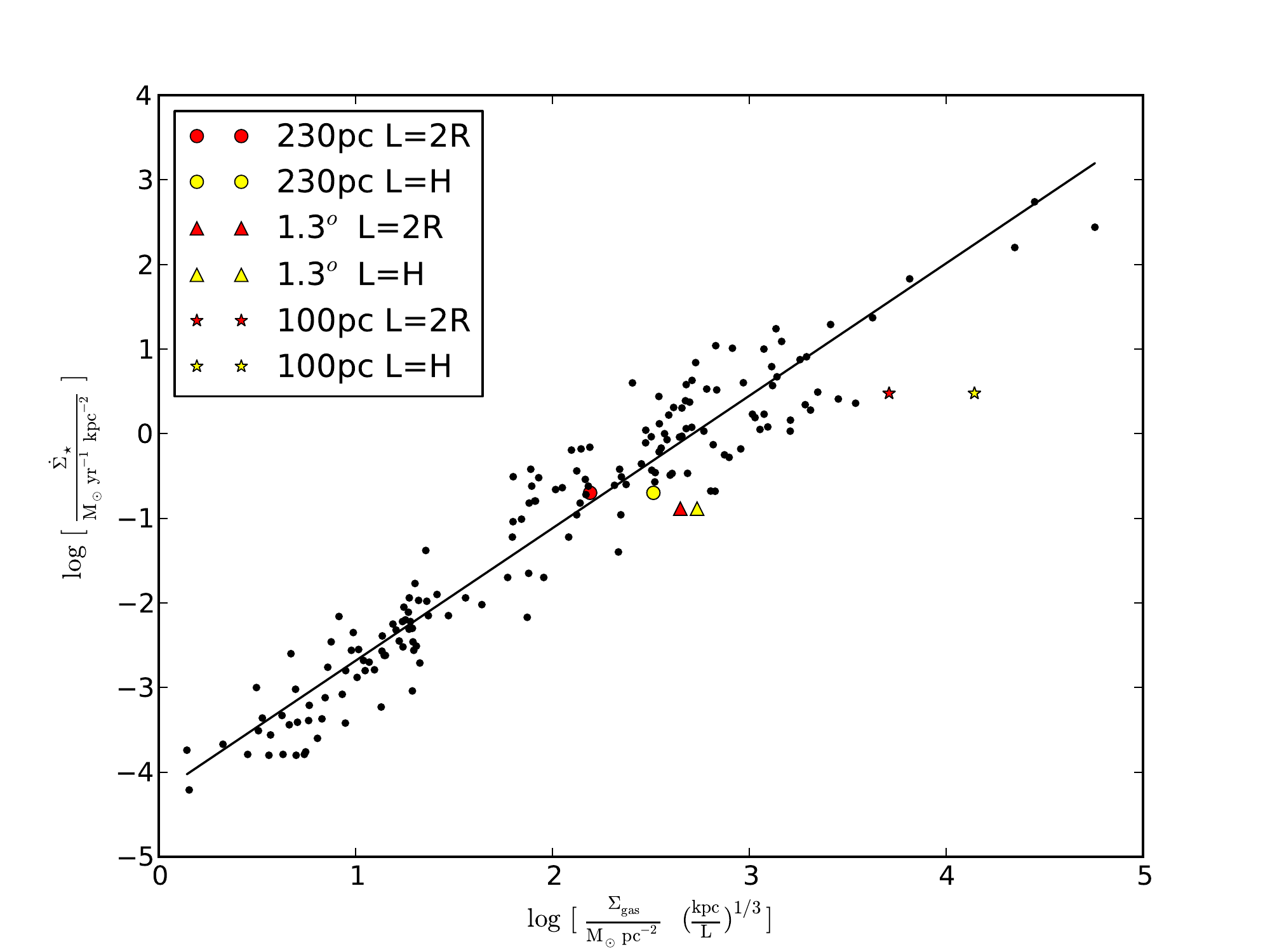}
\caption{SFR density as a function of the gas surface density, divided by the 1/3 power  of the integration scale length L. The yellow symbols  display  the location of three CMZ regions,  using L equals to the  vertical thickness and the red ones, the same regions but with the   disk size as integration scale L. The black circles are the same galaxy/regions displayed in Figure \ref{F2}. }
\label{F4}
\end{center}
\end{figure}

In Figure \ref{F4} we study the location of three regions  of the CMZ in the same star formation plot showed in Figure \ref{F2} (displayed in filled black circles). The regions studied are the central 100pc ring (stars), the $\rm 1.3^{o}$ complex (triangles) and the 230pc zone (circles) that includes the two previous regions, with data taken from Kruijssen et al (2014).  The yellow symbols in Figure \ref{F4} display  the location of three CMZ regions,  using L equals to the  vertical thickness H and the red ones, the same regions but with the   disk size as integration scale L. 

We find that in all cases,  the red circles (L=disk size) are closer to the law given by Eq. \ref{fit} (black line), as expected for a edge-on system like the CMZ. In the particular case of the  central 100pc ring, that was a clear outlier with the previous estimation (yellow star), the more meaningful  choice for  integration scale (L=disk size)  is able to bring this region into the relation (within the scatter). This suggests that a systematic study is needed,  to quantify how much of the scatter seen in the relation is due to projection effects in the estimation of L, however,  this study is beyond the scope of this paper.
 
Finally, we what to point out that the integrated 230pc zone is clearly the closest to the relation given by Eq. \ref{fit} (black line). This is evidence  that filling factors should play a role in the normalization of the law, because its location is far from the average of the other two individual regions included, suggesting  that dilution is playing an important  role. This region with size $\sim$0.5 kpc, start to be representative of  the typical filling factors  at galactic scales and reflects, the differences with filling factors in "individual" star forming regions/clouds and it's effect on the relation.

Alternatively, if the differences between the integrated region and the two individual ones goes beyond differences in filling factors, new physical parameters that affects the SF efficiency may explain the departures of the central 100pc ring and the $\rm 1.3^{o}$ complex from relation given by Eq. \ref{fit} (black line), as we will discuss within the following section.

\section{On the physics  controlling   the characteristic length}

The functional dependence found in Figure \ref{F2} (Eq. \ref{fit}), relies  on the integration scale L. For that reason, is relevant to search for the galactic properties that determines such integration scale. Discarding projection effects and avoiding  edge-on systems, is basically a question about what determines the vertical thickness of such star forming galaxy/region and we will focus the discussion in such a case.

\subsection{The largest scale not stabilized by rotation}

In disk galaxies, the vertical thickness is of the order of the largest scale not stabilized by rotation (Spitzer 1978), $\rm \lambda _{\mathrm{rot}\,} \propto G\Sigma_{gas}/\kappa^{2}$, where $\rm \kappa$    is the epicyclic frequency (see Binney and Tremaine 2008 for a formal expression). The $\rm \lambda _{\mathrm{rot}\,}$ length, is the only scale intermediate between stars and galaxies that has a clear physical basis and  determines the  most massive clumps that are able to collapse (Escala \& Larson 2008). This scale is also relevant  for starburst galaxies, since the bulk of the star formation comes from a massive nuclear disk (Downes \& Solomon 1998). Moreover, even for systems without a large scale ordered motion, like non-coplanar orbiting streams in a merger remnant, this scale can be generalized and is responsible for  the width of individual star-forming  streams (Escala et al 2013). Therefore, the largest scale not stabilized by rotation is a natural candidate for controlling the characteristic integration length for  galaxies. 

Replacing L by $\rm \lambda _{\mathrm{rot}\,}$  in Eq \ref{elaw1} and noticing that  $\rm \kappa$ is aproximatelly equals to $\rm \Omega$, within factors of 2 (Binney and Tremaine 2008), the star formation law have the following expression 

\begin{equation}
\rm  \dot{\Sigma}_{\star} = \epsilon \, \Sigma_{gas} \Omega =  \epsilon \,  \frac{\Sigma_{gas}}{t_{orb}}\,
\label{}
\end{equation} 
which is one of the  formulations studied in Kennicutt (1998) and is consistent with a single star formation law for galaxies up to high-z  (Daddi et al 2010b). 

The second dimensionless parameter depends also on L and when is replaced by $\rm \lambda _{\mathrm{rot}\,}$, is $\rm \Pi_2 = v_{turb}\kappa/G\Sigma_{gas}$, that is easy recognizable as the $\rm 'turbulent'$  version of the Toomre  parameter ($\rm Q_{turb}$; Toomre 1964). Under this scenario, to be in agreement with  the exponential dependence of $\rm \epsilon(\Pi_2)$ seen by Padoan et al. (2012), the star formation law should be  $\rm  \dot{\Sigma}_{\star} = \epsilon \, e^{-Q_{turb}/q}  \Sigma_{gas} \Omega$, being q geometrical factors.  The star formation timescale, $\rm \tau_{SF}= \Sigma_{gas}/\dot{\Sigma}_{\star}$, should also be proportional to  $\rm exp(Q_{turb}/q)\,t_{orb}$. In fact, this exponential  dependence in  $\rm \tau_{SF}$ has been observed in numerical experiments of star formation in galactic disks, finding $\rm \tau_{SF} \propto exp(Q_{turb}/0.61)$ (Li, Mac Low \& Klessen 2005). This numerical simulation do not found the  $\tau_{SF} \propto t_{orb}$ dependence, simply because the orbital time is not varied between simulations (see  their Table 1). Moreover, $\rm \Pi_2 (\lambda _{rot})$ shows the equivalence between results in galactic scale simulations (Li, Mac Low \& Klessen 2005) and MHD simulations of GMCs (Padoan et al. 2012).

Finally, since it is  observed  that $\rm Q_{turb} \sim 1$ for most galaxies, this reconcile the apparent conflict between  the strong exponential dependence on $\rm Q_{turb}$ and star formation laws like KS, that  besides of being independent of $\rm Q_{turb}$, are consistent with observations. Nevertheless, as discussed in the previous section, high-z galaxies as sub-population  tends to show a weaker slope  than 3/2 (Fig \ref{F3}). One possible scenario, is that this weaker slope is a sign of  changes in the efficiency $\epsilon$, due to significant variations  of $\rm Q_{turb}$ in this extreme systems. For example, galaxies undergoing a extreme starburst episode that generates enough turbulence to be in a state of $\rm Q_{turb} > 1$, will produce a  significant decrease of  the efficiency $\epsilon$.

\subsection{The turbulent Jeans scale}

Besides the several positive implications  of identifying $\rm \lambda _{\mathrm{rot}\,}$ as the characteristic length L, still is worthy to explore alternatives. As discussed, from gravitational instability, is clear that $\rm \lambda _{\mathrm{rot}\,}$  (Escala \& Larson 2008) is the characteristic length of collapsing clumps and for unstable disks, such scale is similar to the vertical thickness  because is when rotation starts to stabilize (and support) the system. However, is more often  found in the literature that the characteristic length of collapsing clumps is determined the turbulent jeans scale (e.g. Elmegreen 2002, Kim \& Ostriker 2002). It is important to note, that the condition  $\rm Q_{turb} \sim 1$ is equivalent to have $\rm \lambda _{rot} \sim \lambda _{jeans}$ (Escala \& Larson 2008) and therefore, is hard to distinguish between both scales in most galaxies.

If we instead replace L by the two dimensional $\rm 'turbulent'$ Jeans scale, $\rm \lambda _{jeans} \propto v_{turb}^{2}/G\Sigma_{gas}$, Eq 3 can be rewritten as $\rm  \dot{\Sigma}_{\star} = \epsilon G \Sigma_{gas}^{2} v_{turb}^{-1}$. This  is the same star formation law found in  a scenario where turbulence from a starburst regulates the vertical height  (Eq. 21 in Ostriker \& Shetty 2011). The second dimensionless parameter in such a case, $\rm \Pi_{2} =  v_{turb} (G \Sigma_{gas} \lambda _{jeans})^{-1/2}$, is equals to 1, meaning that $\rm  Q_{turb}$ becomes irrelevant and that this scenario is only valid in the particular case of $\rm Q_{turb} \sim 1$. This also supports the scenario, in which  the vertical height is controlled by large scale galactic potential ($\rm \lambda _{rot}$) and is the self-regulation loop of Goldreich and Lynden Bell (1965), what pushes $\rm \lambda _{jeans} $ towards $\rm \lambda _{rot}$ ($\rm Q_{turb} \sim 1$), as discussed in Escala (2011). 

\subsection{Individual star forming clouds}

On individual star-forming clouds with  sizes smaller  than the galactic vertical thickness, their characteristic length L is their diameter 2R. In order to include the possibility  out of equilibrium configurations, as in \S 4.1, is convenient to express it in terms of the virial parameter $\rm \alpha_{virial} \equiv 2T/W$. Within geometrical factors,  virial parameter is of the order of $\rm Rv_{turb}^2 /GM_{c}$, being $\rm M_{c}=\Sigma_{gas}R^2$, the cloud's total gas mass, or equivalently, $\rm R = v_{turb}^2 /G\Sigma_{gas}\alpha_{virial} $.

Identifiying R as characteristic length in the second dimensionless parameter,  it becomes $\rm \Pi_2 =\alpha_{virial}$. Motivated by the fitted star formation efficiency $\rm \epsilon_{ff} \propto exp(-1.74 v_{turb} (G \Sigma_{gas} L)^{-1/2})$ of Padoan et a. (2012), for individual clouds Eq \ref{elaw3}  can be rewritten as

\begin{equation}
\rm  \dot{\Sigma}_{\star} = \epsilon[\mathcal{M}_s,\mathcal{M}_a] \sqrt{\frac{G}{2R}}\Sigma_{gas}^{3/2} \,\,e^{-1.23\alpha_{virial}}
\label{claw}
\end{equation}



\section{Discussion}

In this paper, we have explored an alternative  approach to study the universal star formation law. Instead of using idealized analytical models to study this  inherently complex and multi-parametric problem, like most in modern astronomy, we used the  Pi theorem of dimensional analysis to search for the relevant physical variables. In addition, this approach avoids  the temptation  for over interpreting  simple $\rm 'spherical \,cow'$ models.

Using the Vaschy-Buckingham Pi theorem, we find that  the star formation law should have a form $\rm  \dot{\Sigma}_{\star} = \epsilon \sqrt{\frac{G}{L}}\Sigma_{gas}^{3/2}$, where L is  a characteristic length. We argued that L should be related with the integration scale, that transforms relevant 3-D quantities in the star formation problem into 2-D ones, like $\rm  \dot{\Sigma}_{\star}$ and $\rm \Sigma_{gas}$. Using simple estimations for L, we find that galaxies from different types and redshifts, including LSB galaxies, and individual star-forming regions in our galaxy, obey this single star formation law. 

The only free parameter introduced in our analysis is $\rm \eta$, in $\rm L = H = \eta R$, that is in principle, a  possible caveat of the present analysis. However, we choose observed values that varies from 0.02 for spirals/LSB galaxies to 2 for individual clouds, a range  at most a factor of 10 in $\rm \sqrt\eta$. For local starburst and high z disk, we choose a larger value of 0.1, that is in agreement with the measured values in starbursts (Downes \& Solomon 1998) and with the larger thickness predicted (Kroupa 2002) and estimated (Elmegreen \& Elmegreen 2005) for gas-rich high z disks. The only debatable value is the choice of $\rm \eta=0.1$ for high z starbursts  (Genzel et al. 2010). However, a choice of $\rm \eta=0.4$, will only increase the scatter in Fig \ref{F2} to  0.45 dex, and $\rm \eta=1$, to 0.47 dex. We decide to do not vary the parameter $\rm \eta$ at this level, since these are variations comparable to other possible sources of error, such as the CO conversion factors assumed in Genzel et al. (2010).

We also find that, depending on the assumption chosen for the vertical scale length H, this   star formation law adopt the    different formulations previously studied in the literature.  For a constant H, we recover the standard KS law.  For $\rm H = \lambda _{\mathrm{rot}}$, we recover  $\rm  \dot{\Sigma}_{\star}  \propto \Sigma_{gas}/t_{orb} $, and for $\rm H = \lambda _{Jeans}$, we find $\rm  \dot{\Sigma}_{\star}  \propto  \Sigma_{gas}^{2}/v_{turb}$.

As vertical scale length, we favour  $\rm H = \lambda _{\mathrm{rot}\,}$, because is the characteristic length of the most massive collapsing clumps and for unstable disks,  such scale is comparable to the vertical thickness, because is when rotation starts to stabilize and globally support the system. In such a case, $\rm \Pi_2$ can be identified as  the Toomre parameter $\rm Q_{turb}$, allowing to include systems out of the equilibrium ($\rm Q_{turb} \ne 1$). This case suggests that for galaxies,  a universal  star formation law of the form:
\begin{equation}
\rm  \dot{\Sigma}_{\star} = \epsilon[\mathcal{M}_s,\mathcal{M}_a]\,e^{-\frac{Q_{turb}}{q}} \sqrt{\frac{G}{\lambda _{rot}}} \Sigma_{gas}^{3/2} \, ,
\label{elawf}
\end{equation}
that is expressed in terms of physical variables related to local and global properties. Using $\rm \lambda _{\mathrm{rot}\,} \propto G\Sigma_{gas}/\kappa^{2}$,  this equation is equivalent to
\begin{equation}
\rm  \dot{\Sigma}_{\star} = \epsilon[\mathcal{M}_s,\mathcal{M}_a]\,e^{-\frac{Q_{turb}}{q}}  \Sigma_{gas} \, \kappa\, ,
\label{elawfB}
\end{equation}
where $\rm \kappa$ is aproximatelly equals to $\rm \Omega$, within factors of 2. This  can be generalized to systems without large scale ordered motion (mergers), by replacing $\rm \kappa$ by the modulus of the orbital frequency vector $\rm \vec{\Omega}_0$ (Escala et al 2013), which  depends also on the center of rotation 0. The exponential decay  on the efficiency in Eq.\ref{elawfB} is so far 
only supported by numerical experiments, therefore, it would be interesting to test such
dependence by observations, more specifically,  if this can  take into account for part of the observed scatter in the star formation law. 

In summary, we have shown the advantages to use  Vaschy-Buckingham Pi theorem, to guide the analysis of the results coming from  numerical simulations  and  observations. Future observations of star formation under more extreme environments, complemented with  new numerical experiments with more physics included and a larger dynamical range, could shed light in finding  new physical variables and its functional dependence  in the star formation law.

I  thank  D. Kruijssen, A. Guzman, F. Becerra, P. Coppi, R. Larson and the anonymous referee for very valuable comments.

\end{document}